\documentclass[preprint,12pt]{aastex}
\usepackage{times}
\begin{document}

\bibliographystyle{apj}

\title{A Compact Cluster of Massive Red Galaxies at a Redshift of $1.51$}

\author{
Patrick J. McCarthy\altaffilmark{1}, 
Haojing Yan\altaffilmark{1},
Roberto G. Abraham\altaffilmark{2},
Erin Mentuch\altaffilmark{2},
Karl Glazebrook\altaffilmark{3},
Lin Yan\altaffilmark{4},
Hsiao-Wen Chen\altaffilmark{5},
S. Eric Persson\altaffilmark{1},
Preethi Nair\altaffilmark{2},
Sandra Savaglio\altaffilmark{6},
David Crampton\altaffilmark{7},
Stephanie Juneau\altaffilmark{8}, 
Damien Le Borgne\altaffilmark{9}, 
R. G. Carlberg\altaffilmark{2}
Ronald O. Marzke\altaffilmark{10},
Inger J{\o}rgensen\altaffilmark{11}, 
Kathy Roth\altaffilmark{11}
Richard Murowinski\altaffilmark{7}, 
}

\altaffiltext{1}{
 Observatories of the Carnegie Institution of Washington,
 813 Santa Barbara Street,
 Pasadena, CA 91101
}
\altaffiltext{2}{
Department of Astronomy \& Astrophysics
University of Toronto, 50 St. George St.
Toronto, ON, M5S~3H4
}
\altaffiltext{3}{
Centre for Astrophysics \& Supercomputing
Swinburne University of Technology
Mail \#31, PO Box 218,
Hawthorne, VIC 3122,
Australia
}
\altaffiltext{4}{
Spitzer Science Center,
California Institute of Technology,
MS 100-22, Pasadena, CA 91125
}
\altaffiltext{5}{
Dept. of Astronomy \& Astrophysics, 
University of Chicago 
5640 South Ellis Ave.
Chicago, IL 60637
}
\altaffiltext{6}{Max-Planck-Institut
  f\"ur extraterrestrische Physik, Garching, Germany}

\altaffiltext{7}{Herzberg Institute of Astrophysics, 
National Research Council, 5071 West Saanich Road, 
Victoria, British Columbia, V9E~2E7, Canada} 

\altaffiltext{8}{
  Steward Observatory
  University of Arizona
  Tucson, AZ 85721 }

\altaffiltext{9}{
DSM/DAPNIA/Service d'Astrophysique
CEA/SACLAY
91191 Gif-sur-Yvetta Cedex, France
}

\altaffiltext{10}{
  Dept. of Physics and Astronomy,
  San Francisco State University,
  1600 Holloway Avenue, 
  San Francisco, CA 94132 
}
\altaffiltext{11}{
  Gemini Observatory,
  Hilo, HI 96720
}

\begin{abstract}
 
   We describe a compact cluster of massive red galaxies at $z=1.51$ 
discovered in one of the Gemini Deep Deep Survey (GDDS) fields. Deep imaging 
with Near Infrared Camera and Multi Object Spectrometer (NICMOS) on the 
{\it Hubble} Space Telescope reveals a high density of galaxies with red
optical to near-IR colors surrounding a galaxy with a spectroscopic redshift of
1.51. Mid-IR imaging with Infrared Array Camera (IRAC) on the {\it Spitzer}
Space telescope shows that these galaxies have spectral energy distributions 
that peak between $3.6\mu$m and $4.5\mu$m. Fits to 12-band photometry reveal 12 
or more galaxies with spectral shapes 
consistent with $z = 1.51$. Most are within $\sim$ 170 co-moving kpc of the GDDS galaxy.
Deep F814W images with the Advanced Camera for Surveys (ACS) on HST reveal that
these galaxies are a mix of early-type galaxies, disk galaxies and close pairs.
The total stellar mass enclosed within a sphere of 170 kpc in radius is 
$> 8 \times 10^{11}M_{\odot}$. The colors of the most massive galaxies are 
close to those expected from passive evolution of simple stellar populations
(SSP) formed at much higher redshifts. We suggest that several of these
galaxies will merge to form a single, very massive galaxy by the present day. This
system may represent an example of a short-lived dense group or cluster core
typical of the progenitors of massive clusters in the present day and
suggests the red sequence was in place in over-dense regions at early times.

\end{abstract}


\section{INTRODUCTION}

\label{sec:introduction}

  Galaxy clusters provide important laboratories for studies of galaxy 
evolution. Key aspects of our knowledge concerning the evolution of massive 
early-type galaxies are derived from the study of rich galaxy clusters at 
intermediate redshifts (e.g. \citet{stanford98,vanDokkum98,kelson00}). Cluster 
studies also provide an important tool for gauging the growth of structure and 
probing the density of the underlying dark matter and energy (e.g. 
\citet{springel05}). Clusters at $z\sim 1 - 2$ provide strong
leverage in addressing both issues and thus there is considerable interest in 
find clusters and groups of galaxies at $z \gtrsim 1$.

  A variety of survey techniques have proven effective at identifying galaxy 
clusters to $z \sim 1$ and beyond. X-ray selected samples at $z > 1$, while 
relatively small, sample the most massive and dynamically relaxed systems (e.g. 
Rosatti et al. 1998; Stanford et al. 1998). Large area multi-color surveys, such as the 
red sequence surveys (e.g. \citet{gladders05}) have produced large samples of 
systems in a range of evolutionary stages. The mid-IR imaging camera on the 
Spitzer Space Telescope (hereafter Spitzer) has extended the range and power of 
multi-color surveys and a number of programs are producing confirmed and 
candidate clusters at $z > 1$ \citep{stanford05,stanford06}. In this short 
report we present the discovery of one of the most distant clusters to date. 
Deep HST and Spitzer images of a massive, passively evolving galaxy identified
in the Gemini Deep Deep Survey (GDDS; \citet{abraham04}) has led to the 
detection of a very compact cluster or group of massive red galaxies at of 
$z=1.51$. Ultra-deep spectroscopy with Gemini and deep multi-band photometry 
has allowed us to obtain a firm redshift for a group of galaxies beyond the
reach of most spectroscopic studies. 

  In \S 2 we present the observations of the galaxy GDDS-12-5869 and its 
surroundings. In section \S 3 we will discuss the properties of the surrounding 
galaxies and in section \S 4 we will briefly discuss the implications for the 
formation of massive galaxies. We adopt the following cosmological parameters: 
$h_o = 0.71$, $\Omega_m = 0.27$ and $\Omega_{\Lambda} = 0.73$. At $z = 1.5$, 
$1^{\prime\prime}$ corresponds to 8.54~kpc.

\section{OBSERVATIONS}

\subsection {Ground-based Imaging and Spectroscopy}

  Much of the basic observational data used in this current work has been 
presented in previous papers related to the GDDS. In paper I \citep{abraham04},
the visible spectroscopy and supporting optical and near-IR imaging data are
described. The optical imaging and early, ground-based near-IR photometry are
discussed in more detail in papers describing the parent sample drawn from the
Las Campanas Infrared survey \citep{chen02,firth02,mccarthy01}. In GDDS 
Paper IV \citep{mccarthy04}, we analyzed spectra of passively evolving galaxies
with $1.3 < z < 2$. One of these objects is GDDS-12-5869, a galaxy at 
$z = 1.51$ with K(Vega)$ = 18.6$ and a moderately red color of $I-K = 2.6$ mag.
Our spectrum of this object is shown in Fig. 1 of Paper IV. In the 
classification scheme of paper I, this object has a high confidence redshift
and a spectrum consistent with no recent or ongoing star formation.

\subsection {HST ACS Imaging}

  Approximately 60\% of the 120 arcmin$^2$ GDDS survey area was imaged with
the Wide Field Camera of ACS on the HST. The details of the observations are
given in paper VIII \citep{abraham07}. The 12-hour GDDS field was observed for
14,640 seconds in the F814W band (hereafter $I_{814}$). The observations were 
dithered in non-integer pixel steps and a mosaic was constructed using the
standard ``MultiDrizzle'' package (Koekemoer et al. 2002). The mosaic is
$\sim 210^{''}\times 210^{''}$ in extent, and GDDS-12-5869 lies 
$35^{''}$ from the southern edge of the field.

\subsection {HST NICMOS Imaging}

\begin{figure}
\plotone{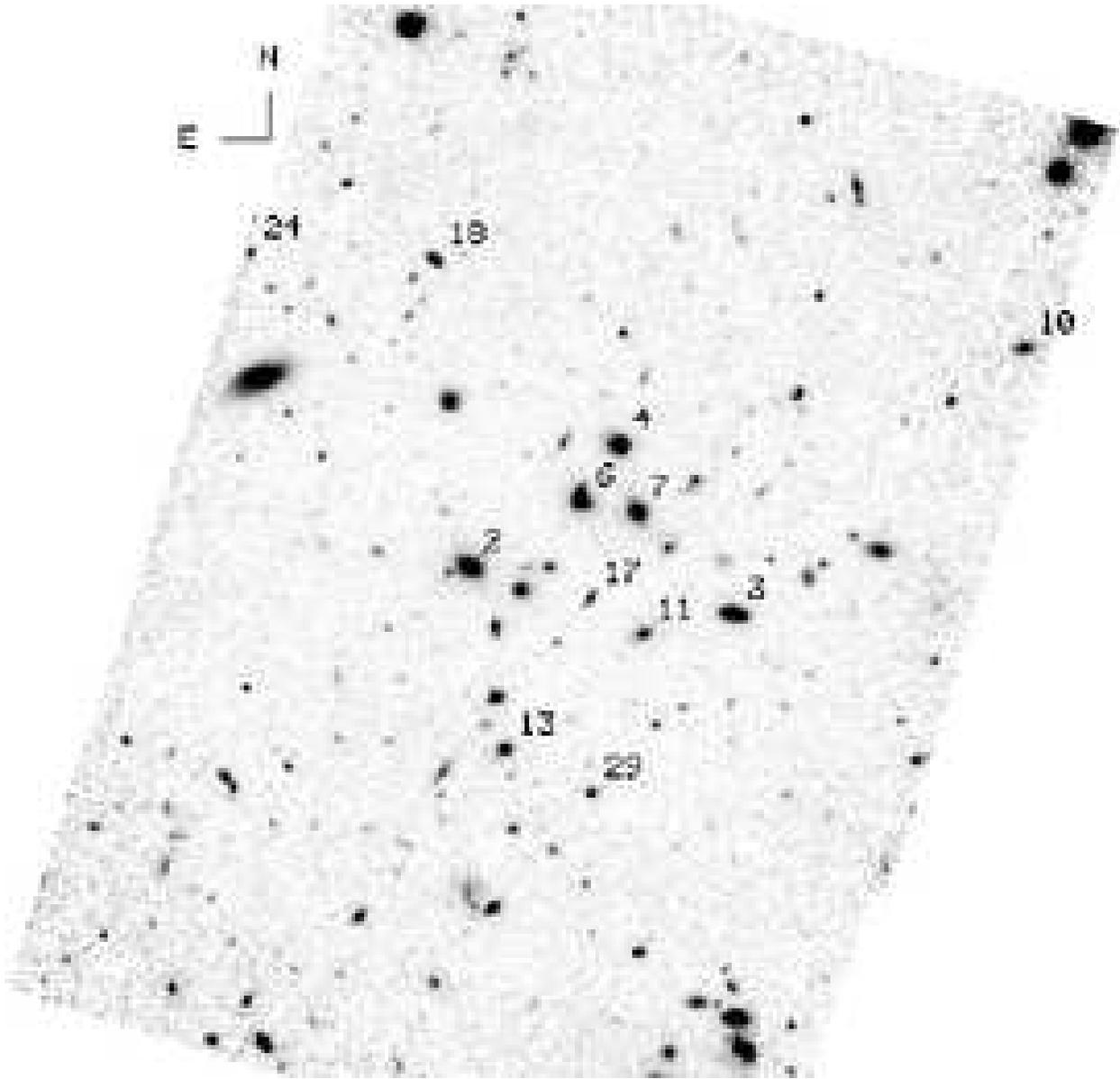}
\caption{\label{fig:nicmos}
NICMOS $H_{160}$ image of the field around GDDS-12-5869 at $z = 1.51$. The
area shown is $51 \times 72$ arcseconds in extent with north at the top
and east to the left. The objects listed in Table 1 are marked. 
GDDS-12-5869 is object \# 4 while GDDS-12-5592 is object \# 13. The objects are 
numbered in order of increasing apparent $H_{160}$ magnitude over the entire NICMOS 
image, only those that we believe to be at $z = 1.5$ are marked.
}
\end{figure}

  Ten passively evolving galaxies with $z>1.5$ drawn from the sample in Paper
IV were observed using NICMOS \citep{thompson92} Camera 3 in the F160W band
(hereafter $H_{160}$).
Each field was observed for a total of 3 orbits, and each orbit was split into
three dithered exposures. The images reduced by the HST on-the-fly-reduction
(OTFR) pipeline were retrieved from the data archive, and were corrected for
various systematics such as pedestal effects, non-linearity, residual 
patterns, etc. These processed images were then combined into mosaics using the
IRAF ``DRIZZLE'' task that implements the ``Drizzle'' algorithm of
\citet{fruchter02}.
Further details of this data set will be provided in a subsequent paper
(McCarthy et al. in prep). 


\subsection {Spitzer IRAC Imaging}

 Three of the GDDS fields were imaged with IRAC \citep{fazio04} as part of a
Guest Observer program.  These fields were observed in all four IRAC channels,
and each field was exposed for 10,842 seconds in each channel. The 
``Basic Calibrated Data'' (BCD) products of our observations were retrieved 
from the {\it Spitzer} data archive, and were further processed to carefully
remove varying backgrounds and residual patterns. These re-processed images then
were combined using the MOPEX package provided by the {\it Spitzer} Science 
Center.

\begin{figure}
\plotone{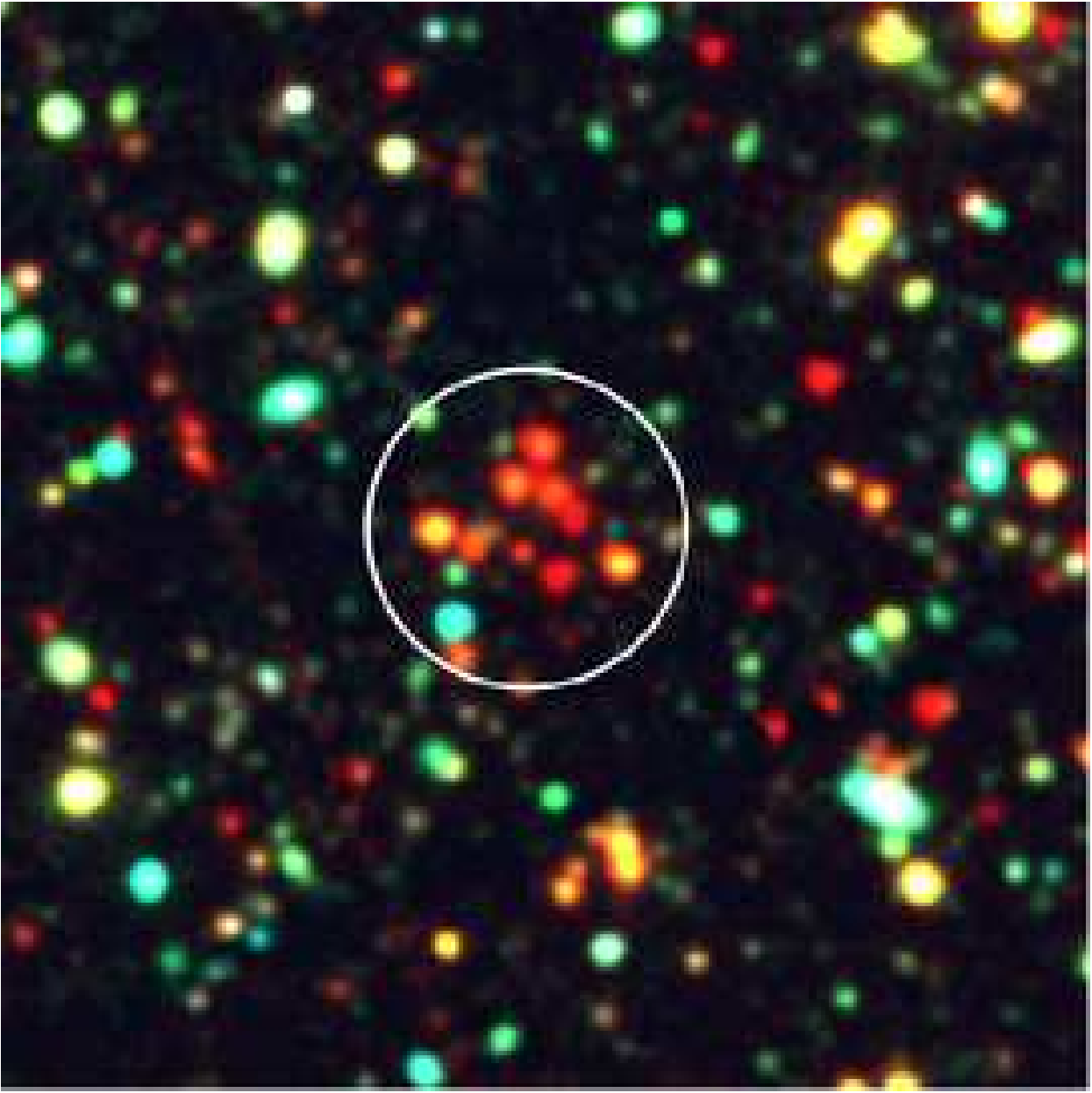}
\caption{\label{fig:colorimages}
Color composite image of the GDDS-12-5869 field. The blue channel is
the sum of the B \& V images, the green channel is the sum of the 
R\& I images, while the red channel combines IRAC channels 1 \& 2 (3.6 and 
4.5$\mu$m). The area shown is $100^{''}\times 100^{''}$ in size. The diameter
of the white circle is 30$^{''}$, corresponding to 256~kpc at $z=1.51$.
}
\end{figure}

\section {RESULTS}

 In Fig. 1 we show the NICMOS image of the field around GDDS-12-5869. In 
addition to the central luminous galaxies, the image reveals a large 
number of faint galaxies concentrated around them. The likely cluster members
are labeled in order of increasing $H_{160}$ magnitude. The surface 
density in a $20^{\prime\prime}$ radius circle to a limiting magnitude of 
$H_{160} = 25$ (AB) in our image is 275~arcmin$^{-2}$. The 
mean surface density of galaxies to this limit in large area random pointings 
is $\sim$ 60~arcmin$^{-2}$ (e.g. Yan et al. 1999).  Thus the NICMOS 
imaging alone reveals a $\sim 5 \times$ overdensity in this field. 

\begin{figure}
\plotone{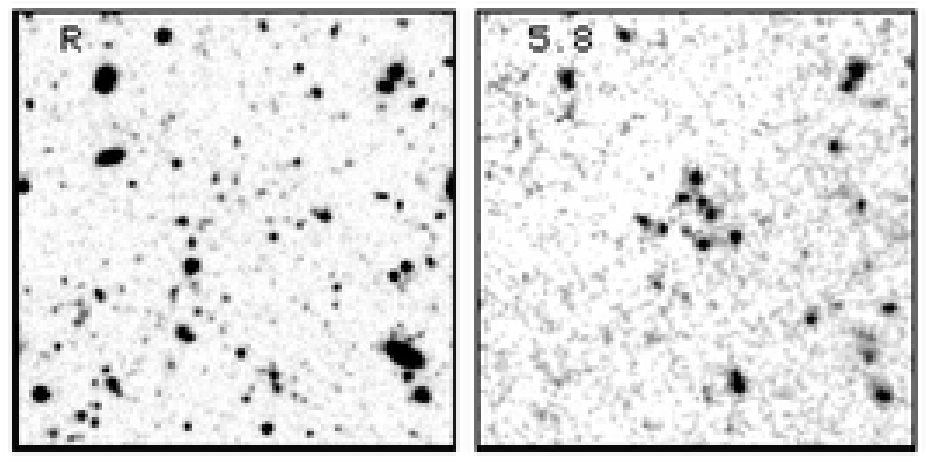}
\caption{\label{fig:panel}
R-band (left) and IRAC channel 3 (right) images of a $2^\prime \times 2^\prime$ 
area around GDDS-12-5869. The concentration of red galaxies is most evident in 
IRAC channel 3 ($5.8\mu$m) as the foreground of galaxies at $z < 1$ drop out. 
}
\end{figure}

\subsection {The Red Galaxies around GDDS-12-5869}

  In Figure 2 we present a color image of this field, which dramatically
illustrates the compact grouping of very red galaxies in the vicinity of 
GDDS-12-5869. There are ten or more galaxies with very red optical-to-near-IR
and optical-to-mid-IR colors within a circle of 128 kpc ($15^{\prime\prime}$)
radius as illustrated by the white circle in Figure 2. As we will show below,
these galaxies all have similar spectral energy distributions (SEDs) and thus
are all likely cluster/group members. A number of other galaxies with similar
colors appear throughout the field, several of which are also consistent with
being at the same redshift.

   The nature of this group of red galaxies is illuminated by the IRAC images
at longer wavelengths. The central group of $\sim 10$ red galaxies are 
clearly detected in channel 3 ($5.8\mu$m), and most are also seen in the 
channel 4 ($8.0\mu$m). This clearly shows that they are at high 
redshift; the foreground of $z < 1$ galaxies drop out of the channel 3 and 4 
images as these bands sample rest-frame wavelengths beyond the $1.6\mu m$ peak
of the SED. This is illustrated in Figure 3 where we show the IRAC channel 3
image along side the R image.  

   In Figure 4, we show a $J-K$ vs. $K$ color-magnitude diagram (CMD) for the
area around GDDS-12-5869. Only galaxies within a 300 kpc ($40^{\prime\prime}$)
radius of the GDDS galaxy are shown. The dashed line is the locus of the Coma
cluster color-magnitude relation \citep{depropris98} transformed to $z = 1.51$
using a model with a simple stellar population (SSP) formed at $z = 5$. The
locus for a non-evolving model is $\sim 0.2$ magnitudes redder. The CMD shows
that the most luminous objects, the central four to five galaxies, have colors
that lie close to the expected cluster red sequence at $z = 1.51$. The bright
galaxies in this CMD are at most a few tenths of a magnitude bluer than the
pure passive evolution line, suggesting that significant star formation ceased
one Gyr or more before the epoch of observation. 

\begin{figure}
\plotone{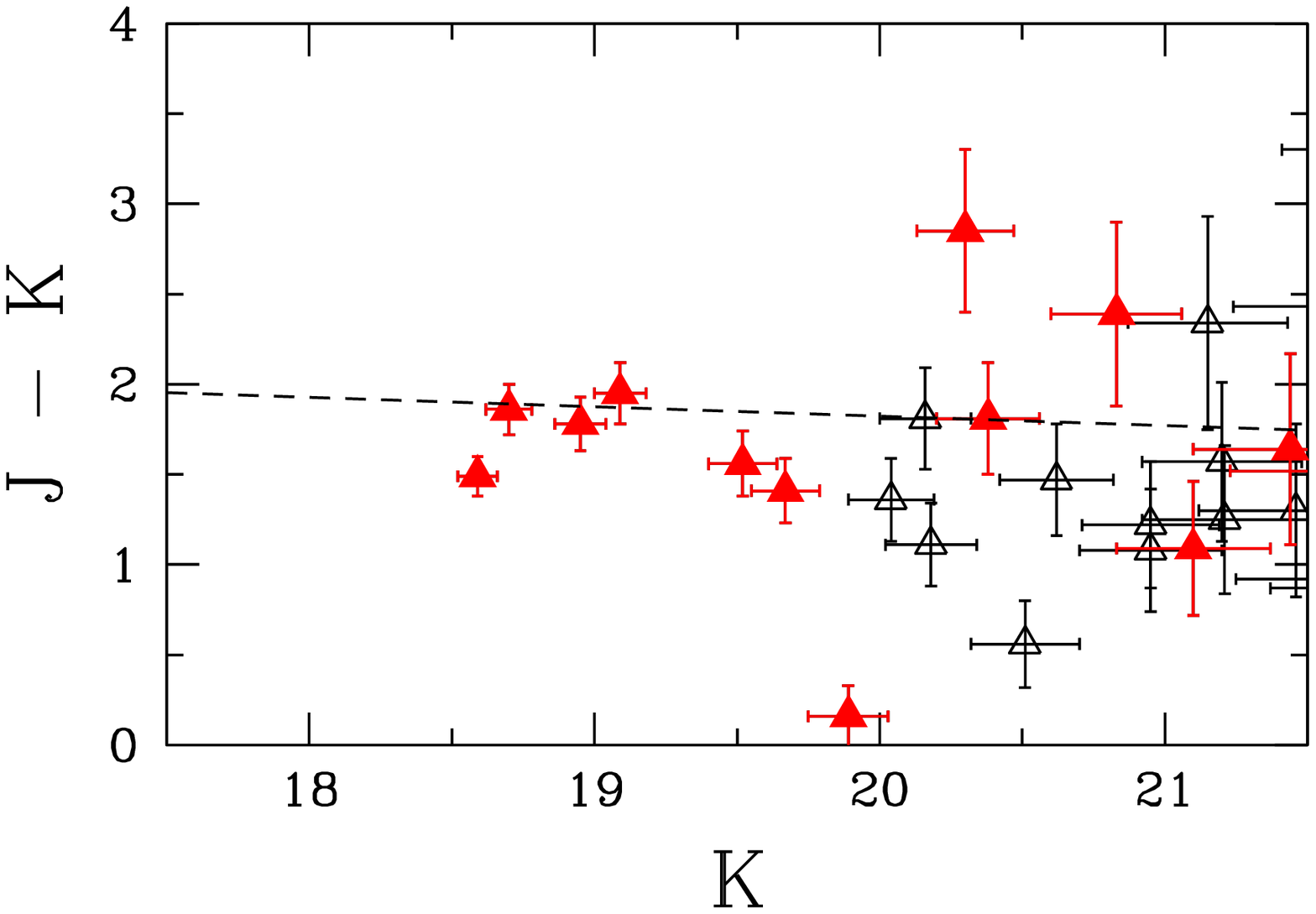}
\caption{\label{fig:cmds}
$J-K$ vs. K(Vega) color-magnitude diagram for a circular region centered on object
\# 17. The red points are objects within a $15^{\prime\prime}$ radius (as in
Figure 1), the black symbols cover an area $30^{\prime\prime}$ in radius.
The dashed line is the color-magnitude relation for the Coma cluster 
evolved to $z = 1.51$, assuming a purely passively evolving SSP formed at a 
redshift of 5.}
\end{figure}

\subsection {Stellar Content of the Red Galaxies}

   In order to further determine which galaxies are likely at the same redshift
as GDDS-12-5869 and to constrain their properties, we fit a range of population
synthesis models of \citet{bruzual03} to the observed SEDs of the galaxies in
this field. The SEDs were constructed using 12-band photometry obtained in 
ground-based $B$, $V$, $R$, $I$, $z^{'}$, $J$, and $K$, NICMOS $H_{160}$, and
IRAC channels 1 to 4.
We adopted a Chabrier initial mass function \citep{chabrier03},
and confined ourselves to single components with solar abundance and 
exponentially declining star formation histories. The stellar mass, age, star
formation time scale, and reddening were left as free parameters. We
first treated the redshift as a free parameter as well, and allowed it to change
at a step-size of $\Delta z=0.1$. The photometric redshifts ($z_{phot}$) thus
derived for the central red galaxies varied somewhat. Two of them (including
GDDS-12-5869) yielded best-fit $z_{phot}=1.4$, consistent with the 
spectroscopic redshift of GDDS-12-5869. The best-fit $z_{phot}$ values for the
others were mixed with some as low as $z_{phot}\sim 0.8$. Nevertheless, a 
redshift of 1.5 was still among the acceptable solutions for most of these
objects. We then fit each of the bright red galaxies again, but forced the
models to be at $z=1.5$. For 12 galaxies we obtained best-fits with
reduced chi-square ($\chi^{'2}$) values near unity. These galaxies are thus
considered as candidate cluster members. Table 1 lists their best-fit model
parameters obtained at a fixed redshift of $z=1.5$, together with their measured
properties.  Several of the galaxies outside the central $20^{\prime\prime}$
radius circle had fits that were significantly worse at $z = 1.5$ than at
sufficiently different redshifts ($|\Delta z|\geq 0.5$), and thus are
unlikely to be cluster members. The total stellar mass in the 7 well measured 
galaxies in the central $30^{\prime\prime}$ is $\sim 6 \times 10^{11}$M$_{\odot}$
and all of these appear to be primarily passively evolving objects at the same
redshift as the GDDS galaxy.

\begin{deluxetable}{ccccccccc}
\tablecaption{Properties and Best Fit Model Parameters of Candidate Cluster 
Members \label{tab:photometry}}
\tablecolumns{14}
\tablewidth{0pc}
\tabletypesize{\small}
\tablehead{
  \colhead{Object} &
  \colhead{RA}      &
  \colhead{DEC}    &
  \colhead{I(Vega)}  &
  \colhead{K(Vega)}          &
  \colhead{m($5.8\mu$)} &
  \colhead{Age(Gyr)} &
  \colhead{M$_{\star} (10^{10} $M$_{\odot}$)} &
  \colhead{$\chi^{'2}$} 
}
\startdata

02  &  12 05 22.31 & -07 24 18.7  & 22.2  & 20.2  & 20.7  & 0.8   & 11.0 & 8.1\\  %
03  &  12 05 21.96 & -07 24 22.3  & 22.7  & 20.3  & 20.6  & 0.9   &  9.7 & 2.6  \\  %
04  &  12 05 21.54 & -07 24 09.5  & 23.8  & 20.5  & 20.5  & 3.5   & 18.8 & 1.1 \\  %
06  &  12 05 21.73 & -07 24 13.6  & 23.3  & 20.7  & 20.9  & 1.4   &  8.2 & 1.0  \\  %
07  &  12 05 21.45 & -07 24 14.6  & 24.0  & 21.0  & 20.8  & 1.9   & 11.4 & 1.0 \\  %
10  &  12 05 19.47 & -07 24 02.1  & 25.2  & 21.8  & 21.1  & 0.5   &  8.0 & 1.1 \\  %
11  &  12 05 21.42 & -07 24 23.8  &  -    & 21.8  & 20.8  & 1.1   &  5.1 & 1.1 \\  %
13  &  12 05 22.13 & -07 24 32.6  & 23.8  & 21.3  & 21.6  & 1.6   &  4.8 & 1.1 \\  %
17  &  12 05 21.69 & -07 24 21.1  & 24.8  &  -    & 22.0  & $<0.1$&  1.8 & 2.3 \\  %
18  &  12 05 22.48 & -07 23 55.4  & 23.9  & 21.8  & 23.1  & 0.6   &  1.7 & 2.3 \\  %
24  &  12 05 23.43 & -07 23 54.9  &   -   &  -    & 21.4  & 0.7   &  1.8 & 14 \\  %
29  &  12 05 21.68 & -07 24 21.7  & 24.5  & 22:   & 22.8  & 1.8   &  1.4  & 0.6\\  %

\enddata
\end{deluxetable}

\begin{figure}
\plotone{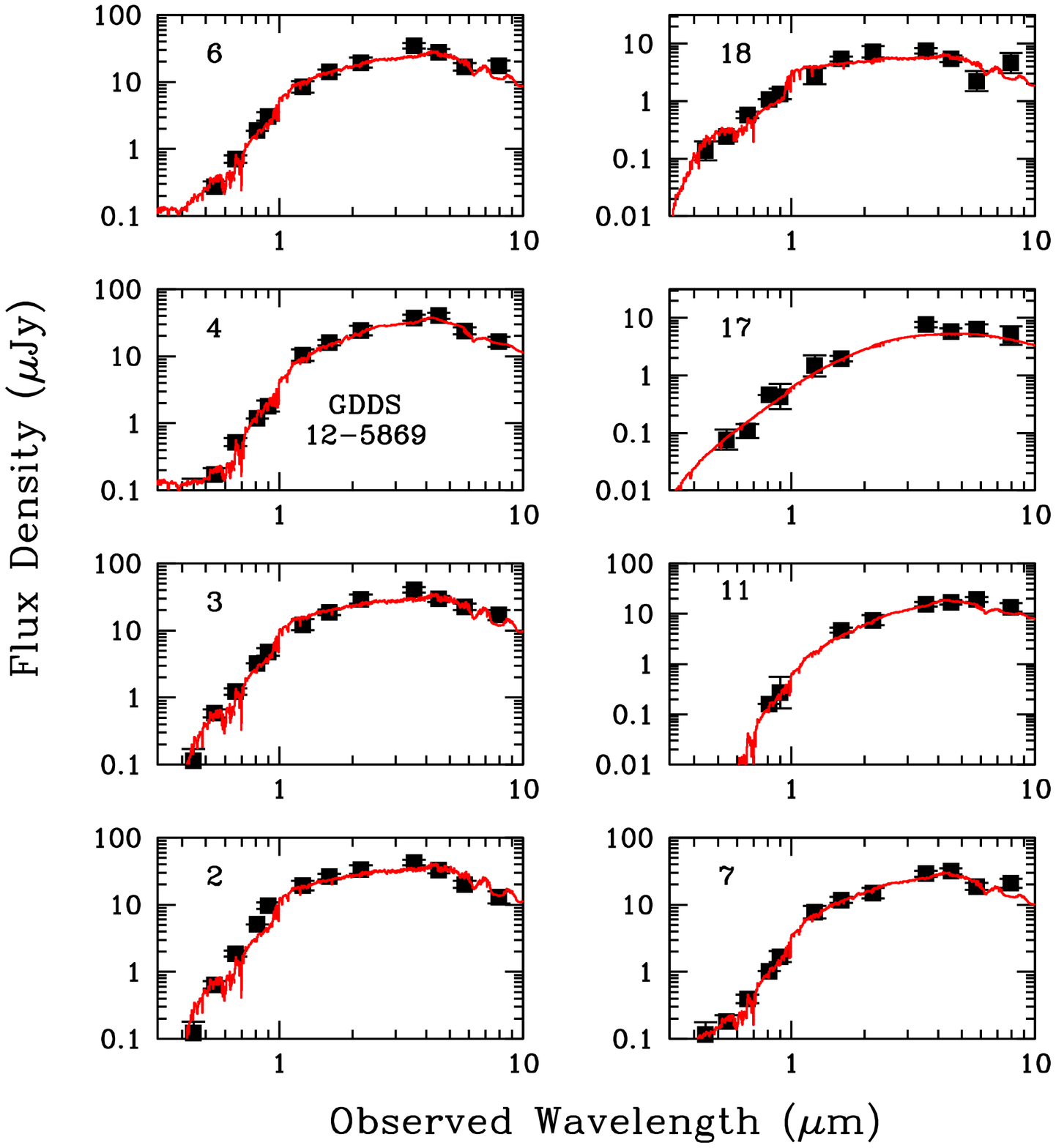}
\caption{\label{fig:seds}
SEDs (filled squares with error bars) and best-fit models at $z=1.5$ (solid
curves) for the eight brightest candidate cluster members. GDDS-12-5869 is
marked.
}
\end{figure}

\subsection {Morphologies}

  Our deep HST imaging allows us to examine the structure of the red galaxies
in this compact group. In Figure 6 we present expanded images of eight 
galaxies from Table 1 in both the $H_{160}$ and $I_{813}$ images from NICMOS and
ACS respectively. Of these eight galaxies, four (Nos. 2, 4, 7, \& 11) appear to
be early-types, while the other four clearly show disks, although galaxies Nos.
3 and 18 appear to have large bulges. Several of the objects have close companions
and galaxy No. 6 appears to be an ongoing merger of two systems. This mix of 
morphologies is not far from that seen in groups at intermediate 
redshifts, but is disk-rich compared to the cores of rich clusters today
\citep{dressler80,whitmore93,zabludoff98}.
Two of the objects in our near-IR images are either not 
detected, or are marginally detected, in our deep ACS $I_{814}$ images. These 
objects (Nos. 11 \& 24) appear to be examples of IRAC-selected extremely red 
objects (IEROs; \citet{yan04}). The object for which we have deep Gemini 
spectroscopy, No. 4 in Figure 1, has the simplest morphology of all of the 
galaxies in the group and appears to be closest to a relaxed early type galaxy.

\begin{figure}
\plotone{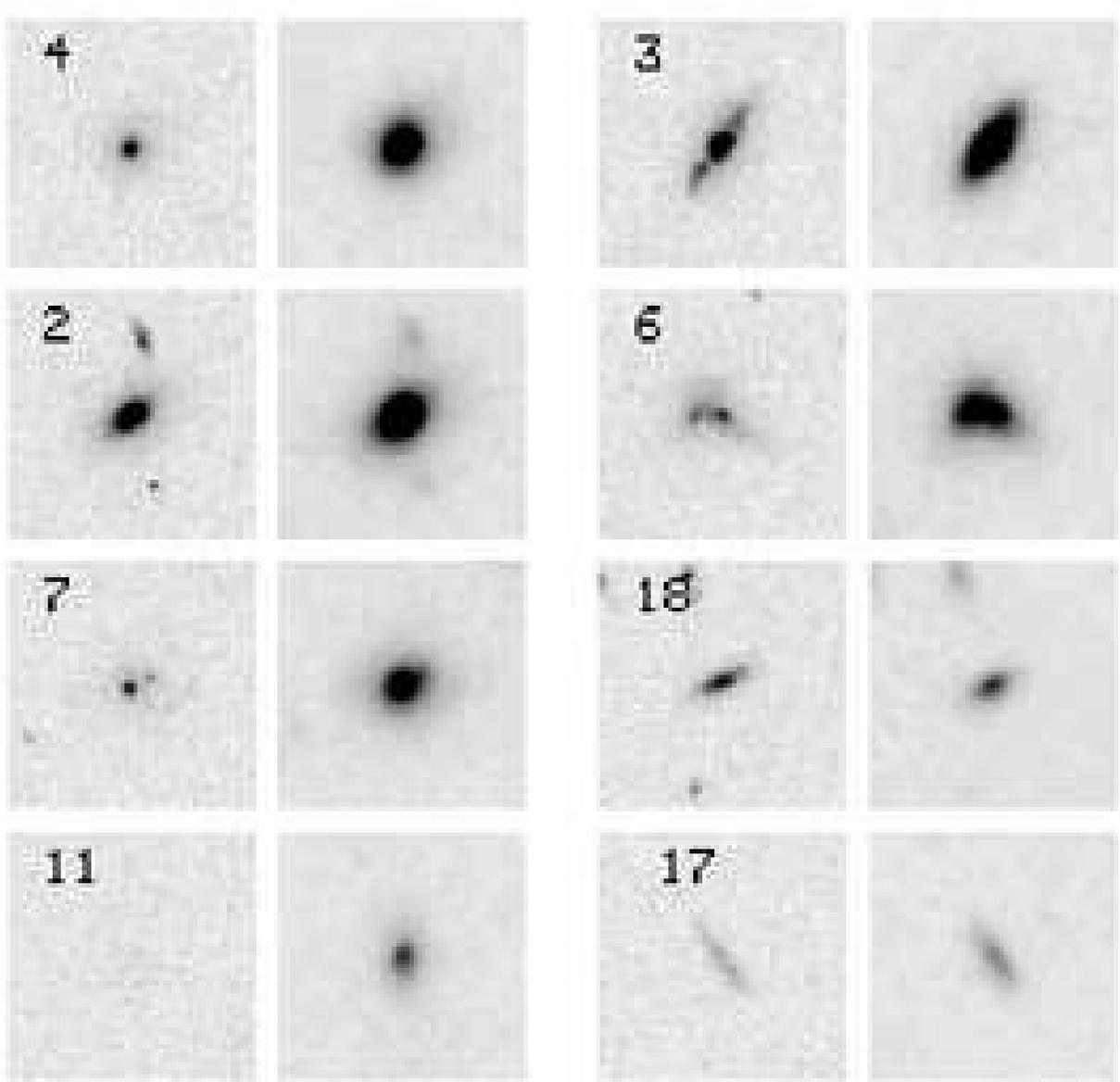}
\caption{\label{fig:cutouts}
NICMOS $H_{160}$ (right) and ACS $I_{814}$ (left) images of cluster candidate
galaxies. The objects are labeled with the same numbers from Figure 1 and Table 1. Galaxies Nos. 4 (GDDS-12-5869), 2, 11 and 7 appear to be early types, while 
3, 17 and 18 are disk galaxies. Object No. 6 appears to be an active merger 
between two disk galaxies.
}
\end{figure}

\section {Discussion}

The cluster around GDDS-12-5869 provides one of the most distant examples of a 
compact red sequence of massive galaxies. This system 
has a crossing time of $\sim 0.5$Gyr if the velocity dispersion is
appropriate for an intermediate mass cluster (e.g. $\sim 500$km s$^{-1}$).
The stellar mass enclosed within a sphere 170~kpc in radius, 
based on our fits, is $\sim 7  - 9\times 10^{11}$M$_{\odot}$, comparable to 
the stellar masses of cD galaxies and to the central density of rich clusters 
(e.g. Henry 1989; Lewis et al. 2003).  It is plausible that several of the 
central galaxies (e.g. nos. 4, 6, \& 7) will merge in a few dynamical times 
and by $z \sim 0.5$ will be a typical brightest cluster galaxy. 
A typical M$_{total}$/M$_{stars}$ for poor clusters (e.g. $\sim 50$;
Balogh et al. 2007) implies a total mass in the central system of 
$\sim 5 \times 10^{13}$M$_{\odot}$.
   
  Dense x-ray selected groups in the local universe have comparable stellar and 
total masses \citep{mulchaey00} and these groups are 
believed to be in a short-lived state of rapid evolution. The compact grouping of 
red galaxies around GDDS-12-5869 may be the core of a larger 
structure. Our deep $H_{160}$ image shows that there are many more faint 
galaxies in this system than detected in the shallower {\it Spitzer} images. The 
overdensity in our $H_{160}$ image and the distribution of colors suggest that the
red sequence may be confined to the central high density region of this system.

The total stellar mass density in the $1 < z < 2$ epoch is evolving rapidly 
\citep{dickinson03,glazebrook04,rudnick03} and there is evidence that the mass 
density in early types in particular is undergoing dramatic change at redshifts 
$\sim 1$ and higher \citep{abraham07}. Deep spectroscopic studies also show 
that pure passive galaxies are becoming increasingly rare at $z \gtrsim 1.5$.
Dense groups and clusters, such as the one  presented here, may be the 
sites of the rapid growth and transformation of high 
mass galaxies that led to the formation of the red sequence seen at $z \sim 1$ 
and below (e.g. Kauffman et al 2003; Faber et al. 2006).

  \end{document}